\documentclass{article}
\usepackage[margin=1in,footskip=0.25in]{geometry}

\usepackage{graphicx}
\usepackage{amsmath}
\usepackage{natbib}
\usepackage[table]{xcolor}
\usepackage{rotating}

\begin{document}

\title{Modeling the Multiple Myeloma Vicious Cycle: Signaling Across the Bone Marrow Microenvironment}

\author{
Catherine E. Patterson\footnote{C. E. Patterson, Program in Applied Mathematical \& Computational Sciences, 14 MacLean Hall, Iowa City, IA 52242, catherine-patterson$@$uiowa.edu} , Bruce P. Ayati\footnote{B. P. Ayati, Department of Mathematics, Program in Applied Mathematical \& Computational Sciences, and Department of Orthopaedics \& Rehabilitation, 14 MacLean Hall, Iowa City, IA 52242} , and Sarah A. Holstein\footnote{S. A. Holstein, Department of Medicine, Roswell Park Cancer Institute, L5-304 CGP, Elm and Carlton Streets, Buffalo, NY 14263}}

\maketitle

\begin{abstract}
Multiple myeloma is a plasma cell cancer that leads to a dysregulated bone remodeling process.  We present a partial differential equation model describing the dynamics of bone remodeling with the presence of myeloma tumor cells.  The model explicitly takes into account the roles of osteoclasts, osteoblasts, precursor cells, stromal cells, osteocytes, and tumor cells.  Previous models based on ordinary differential equations make the simplifying assumption that the bone and tumor cells are adjacent to each other.  However, in actuality, these cell populations are separated by the bone marrow.  Our model takes this separation into account by including the diffusion of chemical factors across the marrow, which can be viewed as communication between the tumor and bone.  Additionally, this model incorporates the growth of the tumor and the diminishing bone mass by utilizing a ``moving boundary.''  We present numerical simulations that qualitatively validate our model's description of the cell population dynamics.
\end{abstract}

\section{Introduction}

Multiple myeloma is a plasma cell cancer characterized by an excess of malignant plasma cells in the bone marrow.  The disease has a significant impact on the bones, the immune system, and the kidneys (\citealt{AmericanCancerSociety}).  In the bone, patients experience pain, hypercalcemia, fractures, and spinal cord compression (\citealt{drake2014}).  Spinal cord compression can lead to severe back pain, numbness, and muscle weakness.  Hypercalcemia, or high levels of calcium in the blood, can result in dehydration, excessive urination, constipation, loss of appetite, weakness, drowsiness, confusion, and even kidney failure or coma.  When the kidneys begin to fail and lose the ability to remove waste from the body, symptoms like weakness, shortness of breath, itching, and leg swelling can arise.  The American Cancer Society expects that in 2015 the United States will see approximately 26,850 new multiple myeloma diagnoses and 11,240 deaths from the disease.  Survival times range from 29 to 62 months once treatment has started  (\citealt{AmericanCancerSociety}).

Some risk factors associated with multiple myeloma include age, gender, race, family history, and obesity.  There are very few myeloma patients under the age of 35 (less than one percent), and most multiple myeloma patients are 65 years of age or older.  Women are less likely to have myeloma than men, and African Americans develop the disease at least twice as often as Caucasian Americans.  While the majority of multiple myeloma patients have no family history of the disease, individuals with a sibling or parent who has had multiple myeloma are four times as likely to have the disease.  Other risk factors include radiation exposure and solitary plasmacytoma (\citealt{AmericanCancerSociety}).

\subsection{Biological Background}

Multiple myeloma bone disease disrupts the body's ability to maintain a healthy skeleton (\citealt{AmericanCancerSociety}).  
Healthy bone continuously remodels itself in order to repair damaged bone, to adapt to mechanical strains, and to gain access to minerals stored in the bone (\citealt{Burr2002, Parfitt2002}).  The bone remodeling process involves the removal of old, and perhaps damaged, bone and its replacement with new bone.  The primary actors in this process are cells called osteoclasts, osteoblasts, and osteocytes. Together, the osteoclasts (which destroy bone) and osteoblasts (which form new bone) are called the basic multicellular unit, or BMU (\citealt{Bellido2014}).

Osteoclasts are responsible for bone removal (also called osteolysis or bone resorption).  
They are multinucleated descendants of the hematopoietic monocyte-macrophage lineage.  
Once the remodeling process has begun, hematopoeitic precursor cells are recruited to the BMU.  
Once there, the precursor cells differentiate into preosteoclasts.  
Then, the mononuclear preosteoclasts join to form the multinucleated mature osteoclast.
These osteoclast precursor cells are recruited from their myeloid progenitors by macrophage colony-stimulating factor (M-CSF), tumor necrosis factors (TNF), interleukin-6 (IL-6), receptor activator of nuclear factor kappa-B ligand (RANKL), growth factors (GFs), and Activin A.
Then mature osteoclast recruitment from the preosteoclast population is regulated by  osteocyte-secreted RANKL and osteoblast secreted osteoprotegerin (OPG).
Once bone resorption is complete, osteoclasts undergo apoptosis.  While the factors that stimulate apoptosis have not yet been completely determined, in vitro experiments have shown that high calcium levels lead to osteoclast apoptosis (\citealt{Bellido2014}).


Osteoblasts are responsible for the creation of new bone; they carry out bone matrix protein secretion and bone mineralization.  Osteoblasts are the descendants of mesenchymal stem cells and are characterized by a cuboidal shape and a large nucleus located at the edge of the cell (\citealt{Bellido2014}).  
As with osteoclasts, osteoblast formation is regulated by chemical factors.  Osteoclast-derived coupling factors recruit osteoblast precursors from a pool of mesenchymal stem cells.  Then the formation of mature osteoblasts is promoted by insulin-like growth factor (IGF), transforming growth factor-$\beta$ (TGF$\beta$), and bone morphogenetic proteins (BMPs) secreted by osteoclasts (\citealt{Parfitt1994, Bonewald1994, Plotkin2014}).  
Once new bone has been formed, $60\%-80\%$ of osteoblasts undergo apoptosis.  Some of the remaining osteoblasts flatten and become lining cells.  The rest become osteocytes (\citealt{Bellido2014, bonewald2011}).


Approximately 5\%-20\% of osteoblasts become trapped in the bone and differentiate into osteocytes.  They are regularly dispersed throughout the mineralized bone and account for over 90\% of the cells in the bone matrix and on the surface of the bone (\citealt{Bellido2014, bonewald2011}).  Osteocytes are located in lacunae and are connected by a network of dendritic processes, which are found in the canaliculi in the bone matrix.  The proteins produced by osteocytes are transported through this network of lacunae and canaliculi.  Thus, osteocytes can influence other cells within the bone matrix and on the surface of the bone (\citealt{buenzli2015}).


Recent studies have also shown that osteocytes play a key role in the regulation of osteoclasts and osteoblasts (\citealt{bonewald2011}).  They are able to identify damaged bone and induce osteoclastogenesis with RANKL (\citealt{Bellido2014, bonewald2011}).  This happens in two ways.  First, osteocytes going through apoptosis cause oseoblasts and stromal cells to produce RANKL, thereby stimulating osteoclast recruitment.  Second, osteocytes can secrete RANKL themselves (\citealt{Bellido2014}).  Osteocytes also produce and secrete sclerostin, which inhibits osteoblast recruitment by blocking the Wnt signaling pathway (\citealt{bonewald2011, neve2012, Kular2012}).

In healthy bone, the destruction of bone by osteoclasts is matched by the replacement of bone by osteoblasts so that bone mass is returned to its original state.  However, in multiple myeloma patients, the bone remodeling process is out of balance. In this unhealthy bone, bone destruction outpaces bone replacement, leaving patients with bone lesions.  These lesions are quite common in multiple myeloma patients; over ninety percent of patients suffer from them.  They occur most often in the spine, skull, pelvis, and ribs.  Bone lesions lead to pathologic fractures, bone pain, hypercalcemia, and spinal cord compression (\citealt{drake2014}).  Even in complete remission, multiple myeloma patients usually do not show reduction of skeletal lesions (\citealt{Wahlin2009}).

Multiple myeloma leads to bone lesions because myeloma tumor cells cause increased osteoclast production, increased osteoclast activity levels, and decreased osteoblast production (\citealt{Mundy1974, Bataille1991, Valentin-Opran1982, Evans1989, Bataille1990}).  This causes increased bone resorption, which in turn encourages tumor growth.  This is called the multiple myeloma ``vicious cycle'' (\citealt{Abe2004}).  This cycle is summarized in Figure \ref{fig:ayatiwiring}.  The details are shown more explicitly in Figure \ref{fig:wiring_withnumbers}.

Myeloma tumor cells encourage this vicious cycle through several chemical factors.  Several of these factors encourage osteoclast production.  Through adhesion betweeen vascular cell adhesion protein 1 (VCAM-1) located on the stromal cells and very late antigen-4 (VLA4) located on the tumor cells, myeloma cells stimulate stromal RANKL production.  This, in turn, simulates osteoclast formation (\citealt{Michigami2000, Mori2004}).
Myeloma cells further encourage osteoclast recruitment through the production of macrophage inflammatory protein-1$\alpha$ (MIP-1$\alpha$), tumor necrosis factor-$\alpha$ (TNF-$\alpha)$, and interleukin-3 (IL-3) (\citealt{Silbermann2013}).  
Myeloma also causes osteocytes to secrete additional interleukin-11 (IL-11), stimulating osteoclastogenesis (\citealt{Giuliani2012}).

Myeloma cells also suppress the recruitment of osteoblasts.  Some chemical factors secreted by myeloma tumor cells that decrease osteoblast production are Dickkopf-related protein 1 (DKK1), IL-3, sclerostin, and secreted frizzled-related proteins (sFRPs) (\citealt{drake2014, Tian2003, Ehrlich2005, Colucci2011, Oshima2005}).  Additionally, tumor cells increase stromal cell production of Activin A, leading to further decreased osteoblast production (\citealt{Vallet2010}).

The other half of the multiple myeloma ``vicious cycle'' is the promotion of tumor growth by osteoclast signaling.  Osteoclasts secrete B-cell activating factor (BAFF) and a proliferation-inducing ligand (APRIL), which lead to increased tumor growth (\citealt{Abe2006}).

\subsection{Mathematical Background}

Power law approximations are a method of representing biological systems pioneered by Savageau for biochemical systems (\citealt{savageau1969, savageau1969a, savageau1970, savageau1976,voit2000}).  They are equations of the form $$\frac{dX_{i}}{dt}=\sum_{j}\gamma_{i}\prod_{k}X_{j}^{g_{ij}},$$ where the $X_{j}$ are the populations present in the biological system and the $\gamma_{i}$ and $g_{ij}$ are parameters that control the growth and decay of the populations.  By expressing the power law instead as $$\frac{dX_{i}}{dt} = \underbrace{\sum_{j}\alpha_{j}\prod_{k}X_{k}^{g_{ik}}}_{\text{growth}} - \underbrace{\sum_{j}\beta_{j}\prod_{k}X_{k}^{h_{ik}}}_{\text{decay}},$$ we separate the equation into two parts: one that promotes growth of the population and another that contributes to decay.  Each part of the equation is the product of a constant ($\alpha_{j}$ or $\beta_{j}$) and the cell populations that contribute to the growth or decay raised to powers ($g_{ik}$ or $h_{ik}$).
This method is used by \citet{komarova2003, Ryser2009, Ryser2010, ayati2010, graham2013}, and it is used in the model we present here.\\

We choose the more qualitatative or phenomenological power law approach over mechanistic models with explicit biochemistry (\citealt{Wang2011, Eudy2015, Ji2015}) for a number of reasons: the models are much simpler mathematically; eventually they will be easier to parameterize from patient data; and the fundamental relationships involved are more robust to changes in the understanding of the underlying biochemistry.   This last point is critical.  A high fidelity mechanistic model, where the parameters are mostly estimated, would indeed provide valuable and quantitatively precise information about the underlying rate constants.   However, if the mechanistic  model is based on assumptions that later turn out to be false, whatever claims that are made about the underlying rate constants will also turn out to be false.  We are operating under the assumption that the current consensus on the mechanisms underlying multiple myeloma bone disease are subject to change. 

\begin{figure}
\begin{center}\includegraphics[width=0.75\textwidth]{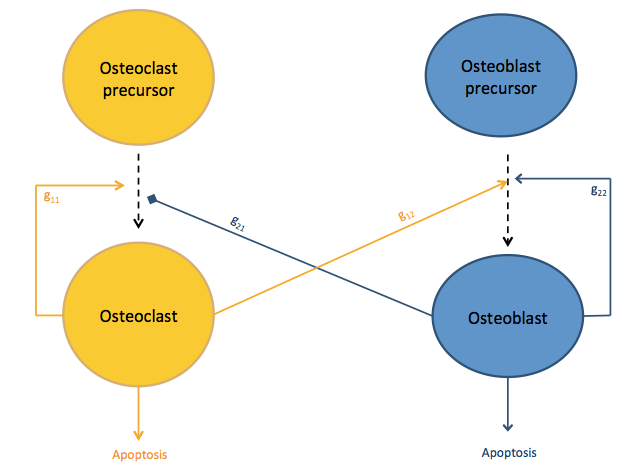}\end{center}
\caption{Diagram of the chemical signals between osteoclasts and osteoblasts, as described by \citet{komarova2003}.  The parameters are also taken from \citet{komarova2003}: $g_{11}$ (autocrine promotion of osteoclasts), $g_{12}$ (paracrine promotion of osteoclasts), $g_{21}$ (paracrine inhibition of osteoclasts), and $g_{22}$ (autocrine promotion of osteoblasts) 
}
\label{fig:komarovawiring}
\end{figure}

\begin{figure}
\begin{center}\includegraphics[width=0.75\textwidth]{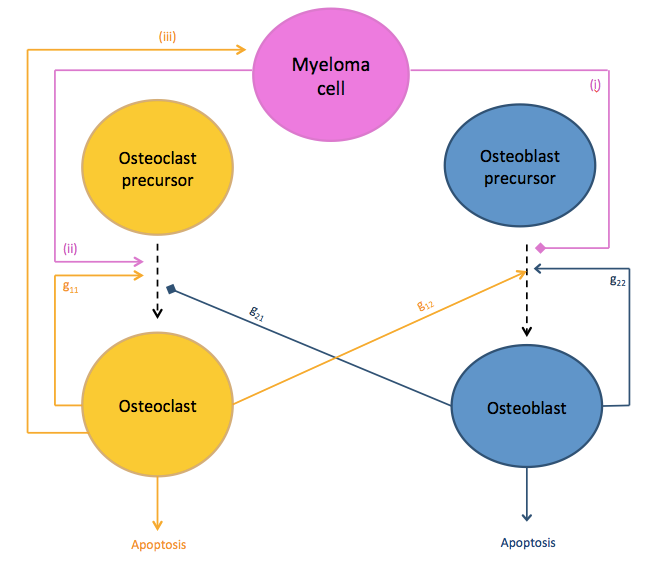} \end{center}
\caption{Diagram of the chemical signals between osteoclasts, osteoblasts, and myeloma tumor cells, as described by \citet{ayati2010}.  The parameters $g_{11}$, $g_{12}$, $g_{21}$, and $g_{22}$ are as in Figure \ref{fig:komarovawiring}.  Arrow (i) represents the suppression of osteoblast production by myeloma tumor cells.  Arrow (ii) represents the increased osteoclast production and activity levels resulting from tumor signaling.  Arrow (iii) represents the increased tumor growth resulting from osteoclast activity.  Together, arrows (ii) and (iii) comprise the multiple myeloma ``vicious cycle'' 
}
\label{fig:ayatiwiring}
\end{figure}

\begin{figure}
\begin{center}\includegraphics[scale=0.6]{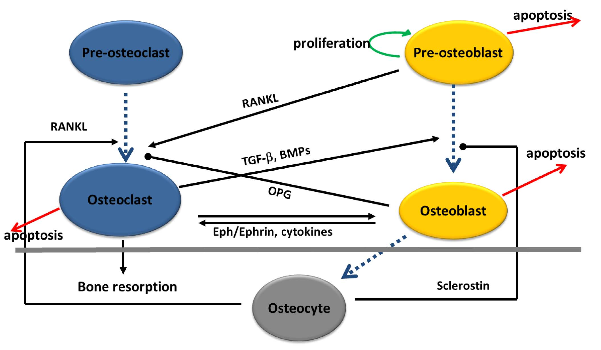} \end{center}
\caption{Wiring Diagram used by \cite{graham2013}}
\label{fig:jasonwiring}
\end{figure}

The model in this paper advances prior work in two main ways.  First, we add a number of additional components we anticipate are necessary if a model is to be able to be used to predict patient outcomes (compare Fig. \ref{fig:wiring_withnumbers} with Figs. \ref{fig:komarovawiring}, \ref{fig:ayatiwiring}, \ref{fig:jasonwiring}).  Second, we have a spatial model that includes cytokine diffusion and and explicit presence of the tumor; the model presented by \citet{Graham2012} used an implicit tumor not located in any particular part of space.  Other models based on ordinary differential equations have no spatial heterogeneity (\citealt{Ryser2009, Ryser2010, Wang2011, Eudy2015, Ji2015}).

\section{Zero-Dimensional Models \label{sec:0dim}}

\begin{figure}
\begin{center}\includegraphics[width=0.75\textwidth]{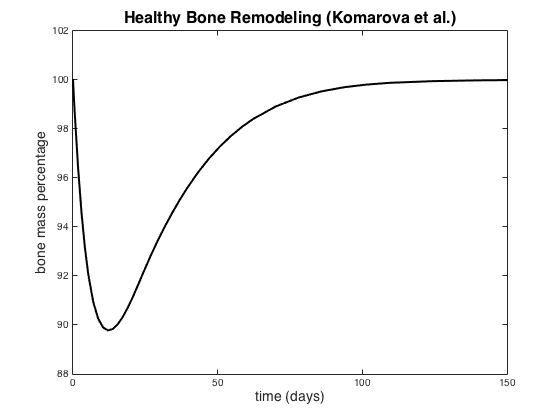} \end{center}
\caption{Simulation of a healthy bone remodeling event (Equations \ref{komarova:c}-\ref{komarova:z}) using the following parameter values: $\alpha_{1}=3$, $\alpha_{2}=4$, $\beta_{1}=0.2$, $\beta_{2}=0.02$, $g_{11}=0.5$, $g_{12}=1$, $g_{21}=-0.5$, $g_{22}=0$, $k_{1}=0.24$, $k_{2}=0.0017$.  The simulation was completed with MATLAB's \texttt{ode15s} with initial conditions $C(0)=15$, $B(0)=316$, and $z(0)=100$ (\citealt{komarova2003}) 
}
\label{fig:bonemasswithouttumor}
\end{figure}

\citet{komarova2003} used Savageau's power law approximations to describe the dynamics of osteoclasts and osteoblasts during a healthy bone remodeling event (without the presence of multiple myeloma tumor cells).  This model takes into account the autocrine and paracrine factors that contribute to the growth and decay of these two cell populations.  The model, based on the cell dynamics described in Figure \ref{fig:komarovawiring}, is
\begin{align}
\label{komarova:c}
\frac{d}{dt}C(t) & = \underbrace{\alpha_{1}
\underbrace{C(t)^{g_{11}}}_{\substack{\text{autocrine}\\\text{promotion}}}
\underbrace{B(t)^{g_{21}}}_{\substack{\text{paracrine}\\\text{inhibition}}}
}_{\text{osteoclast proliferation}} - \underbrace{\beta_{1}C(t),}_{\substack{\text{osteoclast}\\\text{removal}}}\\
\label{komarova:b}
\frac{d}{dt}B(t) & = \underbrace{\alpha_{2}
\underbrace{C(t)^{g_{12}}}_{\substack{\text{paracrine}\\\text{promotion}}}
\underbrace{B(t)^{g_{22}}}_{\substack{\text{autocrine}\\\text{promotion}}}
}_{\text{osteoblast proliferation}} - \underbrace{\beta_{2}B(t),}_{\substack{\text{osteoblast}\\\text{removal}}}\\
\label{komarova:z}
\frac{d}{dt}z(t) & = \underbrace{-k_{1}\max\{0, C(t) - \bar{C}\}}_{\substack{\text{the rate of bone resorption is proportional}\\\text{to the number of osteoclasts}\\\text{exceeding steady-state levels}}} + \underbrace{k_{2}\max\{0, B(t) - \bar{B}\}.}_{\substack{\text{the rate of bone formation is proportional}\\\text{ to the number of osteoblasts}\\\text{exceeding steady-state levels}}}
\end{align}
where $C(t)$ is the density of osteoclasts, $B(t)$ is the density of osteoblasts, and $z$ is the total bone mass.  $\overline{C}$ and $\overline{B}$ represent the steady states for osteoclasts and osteoblasts, respectively.  The steady state is given by
\begin{align*}
\overline{C} & = \left(\frac{\beta_{1}}{\alpha_{1}}\right)^{(1-g_{22})/\Gamma}\left(\frac{\beta_{2}}{\alpha_{2}}\right)^{g_{21}/\Gamma},\\
\overline{B} & = \left(\frac{\beta_{1}}{\alpha_{1}}\right)^{g_{12}/\Gamma}\left(\frac{\beta_{2}}{\alpha_{2}}\right)^{(1-g_{11})/\Gamma},
\end{align*}
where $\Gamma = g_{12}g_{21} - (1-g_{11})(1-g_{22})$.
Figure \ref{fig:bonemasswithouttumor} shows the total bone mass (as a percentage) during a simulation of a bone remodeling event initiated by an increase in osteoclasts. 

\begin{figure}
\begin{center}\includegraphics[width=0.75\textwidth]{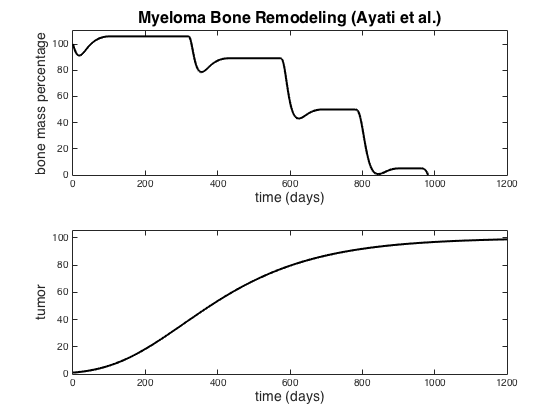} \end{center}
\caption{Simulation of a bone remodeling event with the presence of multiple myeloma tumor cells (Equations \ref{tumor:c}-\ref{tumor:z}) with the following parameter values: $\alpha_{1}=3$, $\alpha_{2}=4$, $\beta_{1}=0.2$, $\beta_{2}=0.02$, $g_{11}=1.1$, $g_{12}=1$, $g_{21}=-0.5$, $g_{22}=0$, $k_{1}=0.0748$, $k_{2}=0.0006395$, $\gamma_{T}=0.005$, $L_{T}=100$, $r_{11}=0.005$, $r_{21}=0$, $r_{12}=0$, and $r_{22}=0.2$.  The simulation was completed with MATLAB's \texttt{ode23t} with initial conditions $C(0)=15$, $B(0)=316$, $z(0)=100$, and $T(0)=1$ (\citealt{ayati2010}) 
}
\label{fig:bonemasswithtumor}
\end{figure}

\citet{ayati2010} expanded on Komarova et al.'s model by including the presence of a multiple myeloma tumor.  The new variables in this model are $T(t)$ (the density of the tumor cells), $L_{T}$ (the maximum tumor size), and $\gamma_{T}$ (the tumor growth constant).  The equations are

\begin{align}
\label{tumor:c}
\frac{d}{dt}C(t) & = \alpha_{1}
\underbrace{C(t)^{g_{11}\left(1+r_{11}\frac{T(t)}{L_{T}}\right)}}_{\substack{\text{increased autocrine}\\\text{promotion of osteoclasts}}}
\underbrace{B(t)^{g_{21}\left(1+r_{21}\frac{T(t)}{L_{T}}\right)}}_{\substack{\text{decreased paracrine}\\\text{inhibition of osteoclasts}}} - \beta_{1}C(t),\\
\label{tumor:b}
\frac{d}{dt}B(t) & = \alpha_{2}
\underbrace{C(t)^{g_{12}/\left(1+r_{12}\frac{T(t)}{L_{T}}\right)}}_{\substack{\text{reduced paracrine}\\\text{promotion of osteoblasts}}}
\underbrace{B(t)^{g_{22}-r_{22}\frac{T(t)}{L_{T}}}}_{\substack{\text{reduced autocrine}\\\text{promotion of osteoblasts}}}
 - \beta_{2}B(t),\\
\label{tumor:t}
\frac{d}{dt}T(t) & \ = \underbrace{\gamma_{T}T(t)\log\left(\frac{L_{T}}{T(t)}\right),}_{\text{Gompertz form}}\\
\label{tumor:z}
\frac{d}{dt}z(t) & = -k_{1}\max\{0, C(t) - \bar{C}\} + k_{2}\max\{0, B(t) - \bar{B}\}.
\end{align}
The parameters $r_{11}, r_{12}, r_{21},$ and $r_{21}$ are all nonnegative.  Thus, the addition of the tumor to this model increases osteoclast production and decreases osteoblast production.
The steady state solution of this model is
\begin{align*}
\overline{C} & = \left(\frac{\beta_{1}}{\alpha_{1}}\right)^{(1-g_{22}+r_{22})/\Lambda}\left(\frac{\beta_{2}}{\alpha_{2}}\right)^{g_{21}(1+r_{21})/\Lambda},\\
\overline{B} & = \left(\frac{\beta_{1}}{\alpha_{1}}\right)^{g_{12}/(\Lambda(1+r_{12}))}\left(\frac{\beta_{2}}{\alpha_{2}}\right)^{(1-g_{11}(1+r_{11}))/\Lambda},\\
\overline{T} & = L_{T},
\end{align*}
where $\Lambda = (g_{12}/(1+r_{12}))(g_{21}(1+r_{21))-(1-g_{11}}(1+r_{11}))(1-g_{22}+r_{22}).$  Computational results for this model are shown in Figure \ref{fig:bonemasswithtumor}.  These results show increasing tumor size accompanied by increased osteoclast activity (bone removal) and decreased osteoblast activity (bone replacement).

\citet{ayati2010} also introduce a model that includes treatment functions.  These treatment functions, $V_{1}(t)$ and $V_{2}(t)$, model the effects of proteasome inhibitors.  Proteasome inhibitors promote osteoblast production and inhibit tumor growth, thereby breaking the multiple myeloma ``vicious cycle.'' The treatment model is

\begin{align}
\label{treatment:c}
\frac{d}{dt}C(t) & = \alpha_{1}C(t)^{g_{11}\left(1+r_{11}\frac{T(t)}{L_{T}}\right)}B(t)^{g_{21}\left(1+r_{21}\frac{T(t)}{L_{T}}\right)} - \beta_{1}C(t),\\
\label{treatment:b}
\frac{d}{dt}B(t) & = \alpha_{2}C(t)^{g_{12}/\left(1+r_{12}\frac{T(t)}{L_{T}}\right)}B(t)^{g_{22}-r_{22}\frac{T(t)}{L_{T}}} - \underbrace{(\beta_{2} - V_{1}(t))}_{\substack{\text{treatment function}\\\text{promotes osteoblast}\\\text{production}}}B(t),\\
\label{treatment:t}
\frac{d}{dt}T(t) & = \underbrace{(\gamma_{T} - V_{2}(t))}_{\substack{\text{treatment function}\\\text{inhibits tumor growth}}}T(t)\log\left(\frac{L_{T}}{T(t)}\right),\\
\label{treatment:z}
\frac{d}{dt}z(t) & = -k_{1}\max\{0, C(t) - \bar{C}\} + k_{2}\max\{0, B(t) - \bar{B}\}.
\end{align}
The treatment functions used in this model are given by
\begin{align*}
V_{1}(t) & = \left\{\begin{array}{ll} 0, & t < t_{\text{start}}\\ v_{1}, & t\geq t_{\text{start}}\end{array}\right.\\
V_{2}(t) & = \left\{\begin{array}{ll} 0, & t < t_{\text{start}}\\ v_{2}, & t\geq t_{\text{start}}.\end{array}\right.
\end{align*}

\begin{figure}
\begin{center}\includegraphics[width=0.75\textwidth]{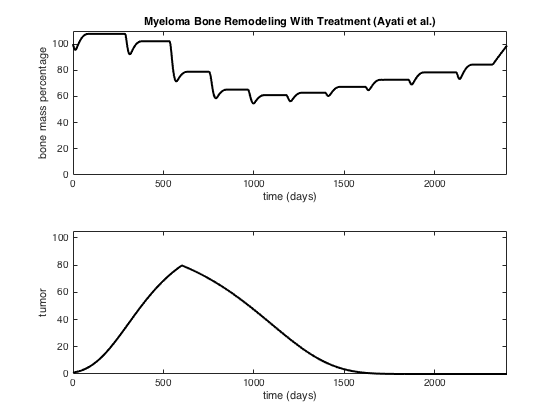} \end{center}
\caption{Simulation of a bone remodeling event with the presence of multiple myeloma tumor cells and treatment (Equations \ref{treatment:c}-\ref{treatment:z}) with the following parameter values: $\alpha_{1}=3$, $\alpha_{2}=4$, $\beta_{1}=0.2$, $\beta_{2}=0.02$, $g_{11}=1.1$, $g_{12}=1$, $g_{21}=-0.5$, $g_{22}=0$, $k_{1}=0.0748$, $k_{2}=0.0006395$, $\gamma_{T}=0.005$, $L_{T}=100$, $r_{11}=0.005$, $r_{21}=0$, $r_{12}=0$, $r_{22}=0.2$, $t_{\text{start}}=600$, $v_{1} = 0.001$, and $v_{2} = 0.008$.  The simulation was completed with MATLAB's \texttt{ode15s} with initial conditions $C(0)=13$, $B(0)=300$, $z(0)=100$, and $T(0)=1$.  The steady states are taken to be $\overline{C}=5$ and $\overline{B}=316$ (\citealt{ayati2010})
}
\label{fig:bonemasswithtreatment}
\end{figure}

Figure \ref{fig:bonemasswithtreatment} shows computational results for this model.  These results are similar to Figure \ref{fig:bonemasswithtumor} until $t=600$, when the treatment is introduced.  At this time, the tumor density begins to shrink.  At the same time, the number of osteoclasts decreases and the number of osteoblasts increases.  This leads to recovery of lost bone mass.

\section{Incorporating Osteocytes \label{sec:jason}}

\citet{graham2013} present a mathematical model of healthy bone remodeling that incorporates two additional cell populations: osteocytes ($Y(t)$) and osteoblast precursors ($B_{P}(t)$).  The biological details of this model are summarized in Figure \ref{fig:jasonwiring}.  The equations for this model are

\begin{align}
\label{jason:y}
\frac{dY}{dt} & = \underbrace{\alpha_{1}B^{g_{31}}\left(1-\frac{Y}{K_{Y}}\right)_{+}}_{\substack{\text{recruitment of osteocytes from osteoblasts}\\\text{that become embedded in the bone}}}\\
\label{jason:bp}
\frac{dB_{P}}{dt} & = \underbrace{\alpha_{2}Y^{g_{21}}\left(1-\frac{Y}{K_{Y}}\right)_{+}^{g_{22}}}_{\substack{\text{differentiation of stem cells}\\\text{to osteoblast precursors}\\\text{by osteocyte signaling}}} + \underbrace{\alpha_{3}B_{P}^{g_{32}}\left(1-\frac{Y}{K_{Y}}\right)_{+}}_{\substack{\text{proliferation of osteoblast}\\\text{precursors by autocrine signaling}\\\text{(when not inhibited by sclerostin)}}} - \underbrace{\beta_{1}B_{P}^{f_{12}}C^{f_{14}}}_{\substack{\text{differentiation of}\\\text{osteoblast precursors to}\\\text{osteoblasts by autocrine}\\\text{and osteoclast signaling}}} - \underbrace{\delta B_{P}}_{\text{apoptosis}}\\
\label{jason:b}
\frac{dB}{dt} & = \underbrace{\beta_{1}B_{P}^{f_{12}}C^{f_{14}}}_{\substack{\text{differentiation of osteoblast}\\\text{precursors into osteoblasts}}} - \underbrace{\beta_{2}B^{f_{23}}}_{\text{apoptosis}} - \underbrace{\alpha_{1}B^{g_{31}}\left(1-\frac{Y}{K_{Y}}\right)_{+}}_{\substack{\text{recruitment of osteocytes from osteoblasts}\\\text{that become embedded in the bone}}}\\
\label{jason:c}
\frac{dC}{dt} & = \underbrace{\alpha_{4}Y^{g_{41}}B_{P}^{g_{42}}(\epsilon + B)^{g_{43}}\left(1-\frac{Y}{K_{Y}}\right)_{+}^{g_{44}}}_{
\substack{\text{differentiation of osteoclast precursors into}\\\text{osteoclasts by the RANK/RANKL/OPG pathway}}} - \underbrace{\beta_{3}C^{f_{34}}}_{\text{apoptosis}}\\
\label{jason:z}
\frac{dz}{dt} & = \underbrace{-k_{1}C}_{\substack{\text{amount of bone removed}\\ \text{is proportional to the} \\\text{number of osteoclasts}}} + \underbrace{k_{2}B}_{\substack{\text{amount of bone formed}\\ \text{is proportional to the} \\\text{number of osteoblasts}}},
\end{align}
where $(x)_{+} = \max\{x,0\}$.

In this model, $K_{Y}$ represents the relationship between osteocyte apoptosis and the decrease in sclerotin inhibition.  The term $\left(1-\frac{Y}{K_{Y}}\right)_{+}$ represents the effects of sclerotin and the Wnt/$\beta$-catenin pathway.  That is, when the number of osteocytes reaches $K_{Y}$, the sclerotin level is sufficient to block Wnt signaling.  This model assumes that osteocyte death is primarily governed by the initiation of the remodeling process.  Thus, no osteocyte apoptosis term is included.

\section{One-Dimensional Bone Remodeling with Multiple Myeloma \label{sec:catie}}

\begin{figure}
\begin{center} \includegraphics[width=0.75\textwidth]{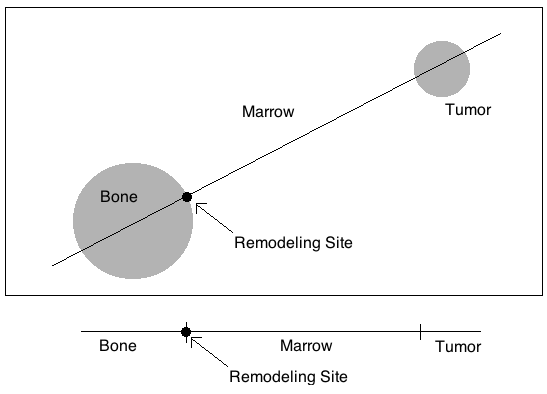} \end{center}
\caption{Diagram of the bone marrow microenvironment.  A section of the bone and a multiple myeloma tumor are separated by the marrow.  A remodeling site (with osteoclasts, osteoblasts, and osteocytes) is located on the edge of the bone}
\label{fig:spatialdiagram}
\end{figure}

Here we present a one-dimensional model of bone remodeling with the presence of multiple myeloma tumor cells.  Figure \ref{fig:spatialdiagram} is a simplified two-dimensional representation of a cross section of a bone marrow biopsy core.  A section of bone and a myeloma tumor lay within the marrow.  Additionally, a remodeling site is located on the edge of the bone.  For our model we consider a one-dimensional representation of this spatial environment, also shown in Figure \ref{fig:spatialdiagram}.

\begin{sidewaysfigure}
\begin{center}\includegraphics[width=\textwidth]{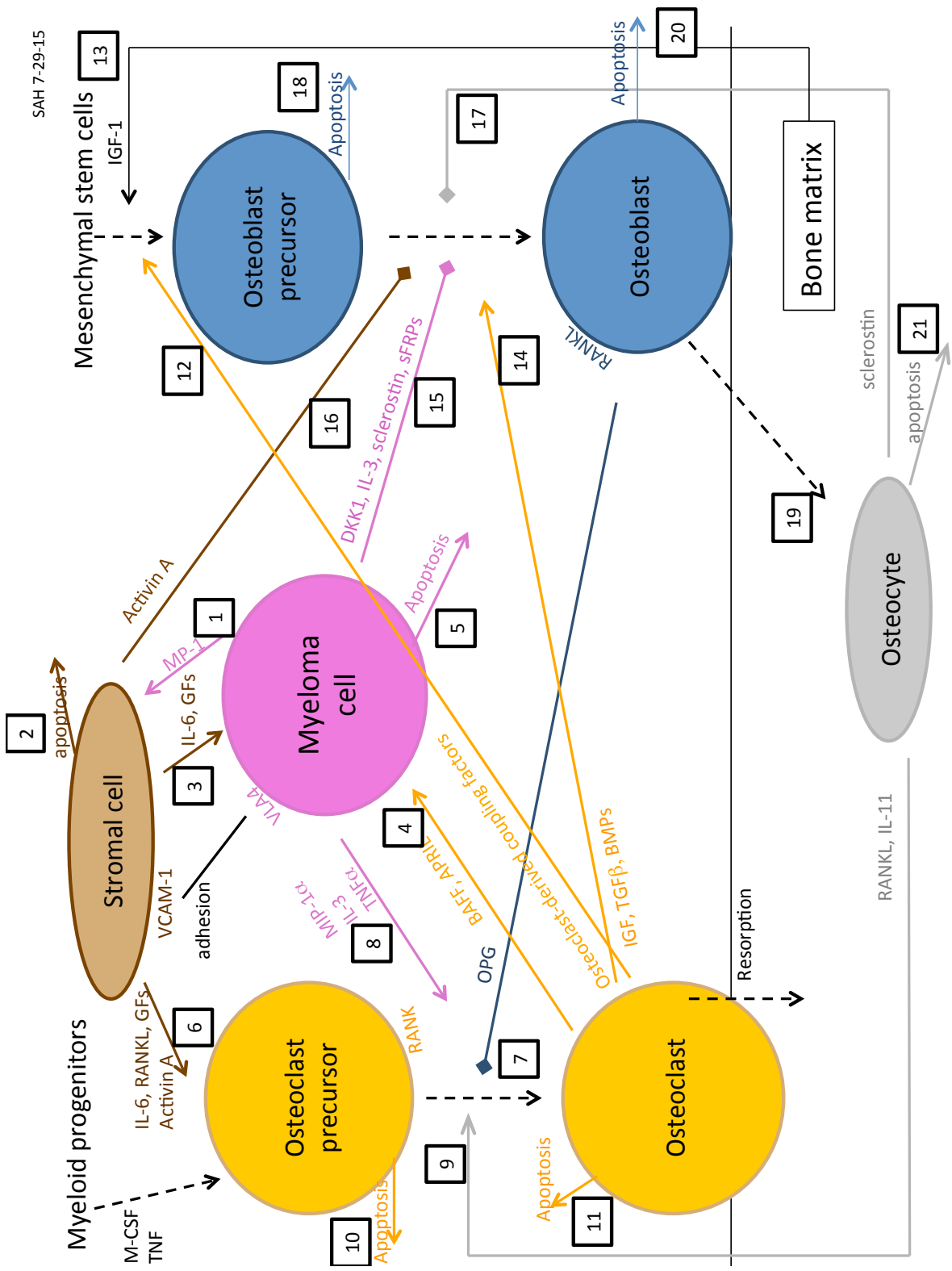}\end{center}
\caption{Wiring diagram used in Equations \ref{eqn:s} - \ref{eqn:r}}
\label{fig:wiring_withnumbers}
\end{sidewaysfigure}

This model builds upon the model presented in \citet{graham2013}  by including additional cell populations, specifically osteoclast precursors ($C_{P}(t)$), stromal cells ($S(t)$), and myeloma tumor cells ($T(t)$).  The interactions of the various cell populations included in this model are detailed in Figure \ref{fig:wiring_withnumbers}.  
This model also incorporates the effects of chemical factors that diffuse across the marrow during the remodeling process:
\begin{itemize}
\item $L_{C}(t)$: BAFF and APRIL, diffusing from the osteoclasts to the tumor cells
\item $L_{T_{1}}(t)$: MIP-1$\alpha$, IL-3, and TNF$\alpha$, diffusing from the tumor cells to the osteoclasts
\item $L_{T_{2}}(t)$: DKK1, IL-3, sclerostin, and sFRPs, diffusing from the tumor cells to the osteoblasts
\item $L_{S_{1}}(t)$: IL-6, RANKL, GFs, and Activin A, diffusing from the stromal cells to the osteoclasts
\item $L_{S_{2}}(t)$: Activin A, diffusing from the stromal cells to the osteoblasts
\end{itemize}
Additionally, the model includes a ``moving boundary.''  That is, the positions of the left and right endpoints of the marrow ($\ell(t)$ and $r(t)$, respectively) are governed by the change in the bone mass and tumor density, respectively.  The equations for this model are

\begin{align}
\label{eqn:s}
\frac{\partial S}{\partial t} & = \underbrace{\alpha_{1}S^{g_{11}}T^{g_{12}}}_{\substack{\framebox{1}\\\text{recruitment of stromal cells}\\\text{by tumor signaling}}} - \underbrace{\beta_{1} S}_{\substack{\framebox{2}\\\text{apoptosis}}} \\
\label{eqn:t}
\frac{dT}{dt} & = \underbrace{\alpha_{2} \left[S^{g_{21}}\right]_{x=r}T^{g_{22}}}_{\substack{\framebox{3}\\\text{recruitment of tumor cells}\\\text{by stromal cell signaling}}} + \underbrace{\alpha_{3} T^{g_{31}}\left[L_{C}^{g_{32}}\right]_{x=r}}_{\substack{\framebox{4}\\\text{recruitment of tumor cells by}\\\text{BAFF and APRIL signaling}}} - \underbrace{\beta_{2}T^{f_{21}}}_{\substack{\framebox{5}\\\text{apoptosis}}} \\
\label{eqn:cp}
\frac{dC_{P}}{dt} & = \underbrace{\alpha_{4} \left[L_{S_{1}}^{g_{41}}\right]_{x=\ell}}_{\substack{\framebox{6}\\\text{recruitment of osteoclast}\\\text{ precursors by stromal}\\\text{cell signaling}}} 
- \underbrace{\gamma_{1}(\epsilon + B)^{h_{11}}\left[L_{T_{1}}^{h_{12}}\right]_{x=\ell}\left(1-\displaystyle\frac{Y}{K_{Y}}\right)_{+}^{h_{13}}Y^{h_{14}}C_{P}^{h_{15}}}_{\substack{\framebox{7}-\framebox{9}\\\text{differentiation of osteoclast precursors into osteoclasts}\\\text{by the RANK/RANKL/OPG pathway}}}  - \underbrace{\beta_{3} C_{P}}_{\substack{\framebox{10}\\\text{apoptosis}}}\\
\label{eqn:c}
\frac{dC}{dt} & = \underbrace{\gamma_{1}(\epsilon + B)^{h_{11}}\left[L_{T_{1}}^{h_{12}}\right]_{x=\ell}\left(1-\displaystyle\frac{Y}{K_{Y}}\right)_{+}^{h_{13}}Y^{h_{14}}C_{P}^{h_{15}}}_{\substack{\framebox{7}-\framebox{9}\\\text{differentiation of osteoclast precursors into osteoclasts}\\\text{by the RANK/RANKL/OPG pathway}}}   - \underbrace{\beta_{4} C}_{\substack{\framebox{11}\\\text{apoptosis}}} \\
\label{eqn:bp}
\frac{dB_{P}}{dt} & = \underbrace{\alpha_{5} C^{g_{51}}}_{\substack{\framebox{12}\\\text{recruitment of}\\\text{osteoblast precursors}\\\text{by osteoclasts}}} + \underbrace{\alpha_{6}z^{g_{52}}}_{\substack{\framebox{13}\\\text{recruitment of osteoblast} \\\text{precursors by IGF-1,} \\\text{secreted by the bone matrix}}}\\ & \hspace*{.5in} -\underbrace{\gamma_{2} C^{h_{21}}\left[L_{T_{2}}^{h_{22}}\right]_{x=\ell}\left[L_{S_{2}}^{h_{23}}\right]_{x=\ell}\left(1-\displaystyle\frac{Y}{K_{Y}}\right)_{+}^{h_{24}}Y^{h_{25}}B_{P}^{h_{26}}}_{\substack{\framebox{14}-\framebox{17}\\\text{differentiation of osteoblast precursors into osteoblasts}}} - \underbrace{\beta_{5} B_{P}}_{\substack{\framebox{18}\\\text{apoptosis}}} \nonumber\\
\label{eqn:b}
\frac{dB}{dt} & = \underbrace{\gamma_{2} C^{h_{21}}\left[L_{T_{2}}^{h_{22}}\right]_{x=\ell}\left[L_{S_{2}}^{h_{23}}\right]_{x=\ell}\left(1-\displaystyle\frac{Y}{K_{Y}}\right)_{+}^{h_{24}}Y^{h_{25}}B_{P}^{h_{26}}}_{\substack{\framebox{14}-\framebox{17}\\\text{differentiation of osteoblast precursors into osteoblasts}}} - \underbrace{\gamma_{3} B^{h_{31}}\left(1-\displaystyle\frac{Y}{K_{Y}}\right)_{+}^{h_{32}}}_{\substack{\framebox{19}\\\text{differentiation of osteoblasts}\\\text{into osteocytes}}} - \underbrace{\beta_{6} B}_{\substack{\framebox{20}\\\text{apoptosis}}}\\
\label{eqn:y}
\frac{dY}{dt} & = \underbrace{\gamma_{3} B^{h_{31}}\left(1-\displaystyle\frac{Y}{K_{Y}}\right)_{+}^{h_{32}}}_{\substack{\framebox{19}\\\text{differentiation of osteoblasts}\\\text{into osteocytes}}} - \underbrace{\beta_{7} Y}_{\substack{\framebox{21}\\\text{apoptosis}}} 
\end{align}
\begin{align}
\label{eqn:lc}
\frac{\partial L_{C}}{\partial t} & = \underbrace{\delta_{11}\nabla^{2}L_{C} - \delta_{12}L_{C}}_{\substack{\framebox{4}\\\text{diffusion of BAFF and APRIL} \\\text{from osteoclasts to tumor cells}}}\\
\label{eqn:lt1}
\frac{\partial L_{T_{1}}}{\partial t} & = \underbrace{\delta_{21}\nabla^{2}L_{T_{1}} - \delta_{22}L_{T_{1}}}_{\substack{\framebox{8}\\\text{diffusion of MIP-1$\alpha$, IL-3, and TNF$\alpha$} \\\text{from tumor cells to osteoclasts}}}\\
\label{eqn:lt2}
\frac{\partial L_{T_{2}}}{\partial t} & = \underbrace{\delta_{31}\nabla^{2}L_{T_{2}} - \delta_{32}L_{T_{2}}}_{\substack{\framebox{14}\\\text{diffusion of DKK1, IL-3, sclerostin,} \\\text{and sFRPs from tumor cells to osteoblasts}}}\\
\label{eqn:ls1}
\frac{\partial L_{S_{1}}}{\partial t} & = \underbrace{\delta_{41}\nabla^{2}L_{S_{1}} - \delta_{42}L_{S_{1}}}_{\substack{\framebox{6}\\\text{diffusion of IL-6, RANKL, GFs}\\\text{and Activin A from stromal cells}\\\text{to osteoclast precursors}}} &\\
\label{eqn:ls2}
\frac{\partial L_{S_{2}}}{\partial t} & = \underbrace{\delta_{51}\nabla^{2}L_{S_{2}} - \delta_{52}L_{S_{2}}}_{\substack{\framebox{15}\\\text{diffusion of Activin A from}\\\text{stromal cells to osteoblasts}}}\\
\label{eqn:z}
\frac{dz}{dt} & = \underbrace{-k_{1}C}_{\substack{\text{amount of bone removed}\\ \text{is proportional to the} \\\text{number of osteoclasts}}} + \underbrace{k_{2}B}_{\substack{\text{amount of bone formed}\\ \text{is proportional to the} \\\text{number of osteoblasts}}} \\
\label{eqn:l}
\frac{d\ell}{dt} &= \underbrace{a \frac{dz}{dt}}_{\substack{\text{movement of the left} \\\text{boundary is proportional to} \\\text{the change in bone mass}}}\\
\label{eqn:r}
\frac{dr}{dt} &= \underbrace{b \frac{dT}{dt}}_{\substack{\text{movement of the right} \\\text{boundary is proportional to} \\\text{the change in the tumor}}}
\end{align}
where $(x)_{+} = \max\{x,0\} = \left\{\begin{array}{ll}x, & x \geq 0 \\ 0, & x<0.\end{array}\right.$
The boxed numbers correspond with the cell signaling represented in Figure \ref{fig:wiring_withnumbers}.

Equation \ref{eqn:s} describes the dynamics of the stromal cell population.  The stromal cells (the connective tissue cells of the bone marrow) are recruited by tumor cell signaling \framebox{1} at a rate $\alpha_{1}$.  Stromal cell apoptosis \framebox{2} occurs at a rate $\beta_{1}$.

Equation \ref{eqn:t} describes the dynamics of the tumor cell population. Myeloma tumor cells are recruited by stromal cell signaling \framebox{3} at the right endpoint of the marrow.  This recruitment occurs at a rate $\alpha_{2}$.  Tumor cells are also recruited by osteoclast signaling of BAFF and APRIL \framebox{4} as a part of the multiple myeloma ``vicious cycle.''  This recruitment occurs at a rate $\alpha_{3}$ and is due to the amount go these ligands present at the right endpoint of the marrow.  Finally, tumor cell apoptosis \framebox{5} occurs at a rate $\beta_{2}$.

The dynamics of the osteoclast precursor cells are described in equation \ref{eqn:cp}.  This equation states that osteoclast precursors descend from a pool of myeloid progenitors \framebox{6} at a rate $\alpha_{4}$.  This differentiation is largely influenced by stromal cell signaling at the left boundary point of the marrow.  Additionally, this equation states that osteoclast precursors differentiate into osteoclasts by the RANK/RANKL/OPG pathway \framebox{7}-\framebox{9} at a rate $\gamma_{1}$.  Finally, we have osteoclast precursor death \framebox{10} at a rate $\beta_{3}$.

Equation \ref{eqn:c} describes the dynamics of the osteoclast population.  This equation states the osteoclasts differentiate from the pool of osteoclast precursors by the RANK/RANKL/OPG pathway \framebox{7}-\framebox{9} at a rate $\gamma_{1}$.  Additionally, osteoclasts undergo apoptosis \framebox{11} at a rate $\beta_{4}$.

Equation \ref{eqn:bp} describes the dynamic of the osteoblast precursor population.  Osteoblast precursors differentiate from a pool of mesenchymal stem cells due to osteoclast \framebox{12} and bone matrix \framebox{13} signaling.  Osteoblast precursors are recruited by osteoclasts at a rate $\alpha_{5}$ and by IGF-1 (secreted by the bone matrix) at a rate $\alpha_{6}$.
Additionally, osteoblast precursors differentiate into mature osteoblasts \framebox{14}-\framebox{17} at a rate $\gamma_{2}$.  Finally, osteoblast precursors undergo apoptosis \framebox{18} at a rate $\beta_{5}$.

The dynamics of mature osteoblasts are described by Equation \ref{eqn:b}.  This equation states that osteoblast precursors are differentiated into osteoblasts \framebox{14}-\framebox{17} at a rate $\gamma_{2}$.  Additionally,  under this model, mature osteoblasts have one of two fates: differentiation into osteocytes \framebox{19} or cell death \framebox{20}.  Osteoblasts differentiate into osteocytes at a rate $\gamma_{3}$ and undergo apoptosis at a rate $\beta_{6}$.

Equation \ref{eqn:y} describes the dynamics of the osteocyte population.  This equation states that osteocytes differentiate from the pool of osteoblasts \framebox{19} at a rate $\gamma_{3}$.  These cells undergo apoptosis \framebox{21} at a rate $\beta_{7}$.

Equations \ref{eqn:lc}-\ref{eqn:ls2} describe the diffusion of chemical factors across the bone marrow.  Each of these equations is a diffusion equation of the form $\frac{dL_{i}}{dt} = \delta_{il}\nabla^{2}L_{i} - \delta_{i2}L_{i}$, where $\nabla^{2}$ is the Laplace operator.

Equation \ref{eqn:z} gives the rate of change of the bone mass.  This equation states that bone resorption is proportional to the number of osteoclasts (with proportionality constant $k_{1}$).  Similarly, bone formation is proportional to the number of osteoblasts (with proportionality constant $k_{2}$).

Equations \ref{eqn:l} and \ref{eqn:r} describe the movement of the bone/marrow interface and the marrow/tumor interface, respectively.  The bone/marrow interface ($\ell(t)$) moves to the left as the bone mass decreases.  Similarly, the marrow/tumor interface ($r(t)$) move to the left as the tumor grows.

\section{Results}

\begin{figure}
\begin{center}\includegraphics[width=0.75\textwidth]{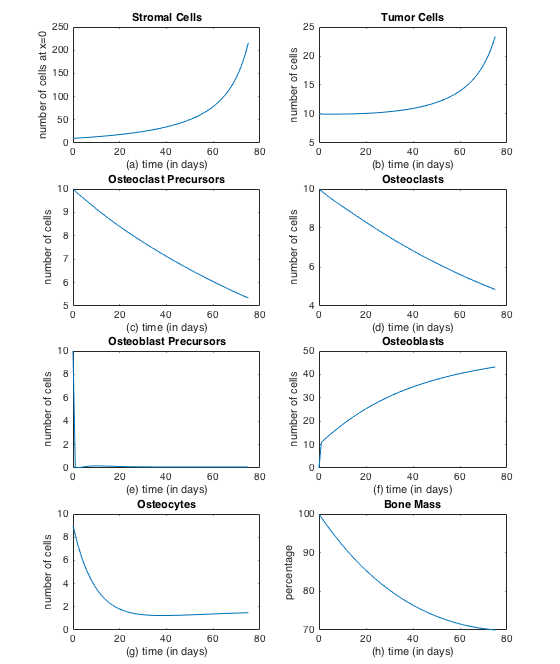}\end{center}
\caption{Computational results for Equations \ref{eqn:s} - \ref{eqn:r}.  This simulation represents a myeloma-dysregulated remodeling event taking place over 75 days}
\label{fig:compresults_cells} 
\end{figure}

\begin{figure}
\begin{center}\includegraphics[width=\textwidth]{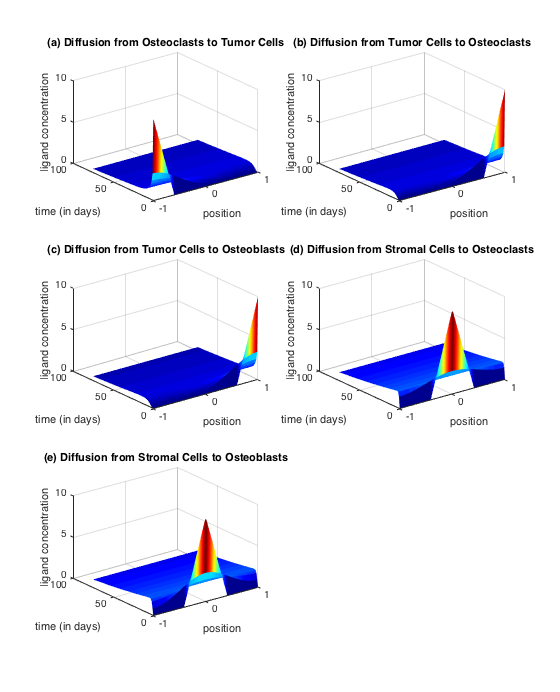}\end{center}
\caption{Computational results for Equations \ref{eqn:s} - \ref{eqn:r}.  This simulation represents a myeloma-dysregulated remodeling event taking place over 75 days}
\label{fig:compresults_diffusion}
\end{figure}

\begin{figure}
\begin{center}\includegraphics[width=\textwidth]{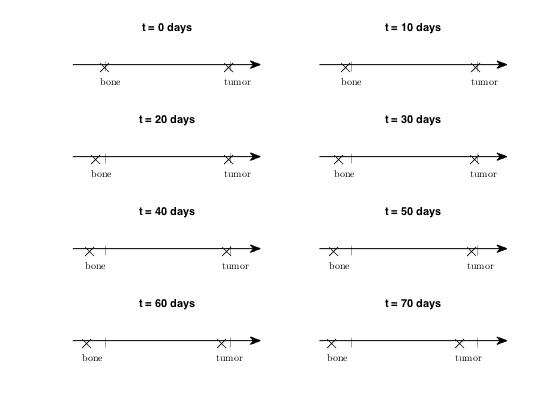}\end{center}
\caption{Computational results for Equations \ref{eqn:s} - \ref{eqn:r}.  This simulation represents a myeloma-dysregulated remodeling event taking place over 75 days}
\label{fig:compresults_boundary}
\end{figure}

Equations \ref{eqn:s} - \ref{eqn:r} were solved using MATLAB's \texttt{pdepe} function  CITE with the parameter and initial condition values listed in Table \ref{tab:parameters3}.  The diffusion values ($\delta_{i1}$) were computed based on the relationship between the size of the peptides (Stokes radius) and the known diffusion values:
\begin{align*}
(\text{Stokes Radius}) & = 0.0156(\text{molecular weight}) + 1.527\\
(\text{diffusion constant}) & = -4 \times 10^{-7} (\text{Stokes Radius}) + 2 \times 10^{-6}
\end{align*} The computed values for each ligand are given in Table \ref{tab:diffusion}  The simulation represents a myeloma-dysregulated bone remodeling event taking place over 75 days.  The results are shown in Figures \ref{fig:compresults_cells}, \ref{fig:compresults_diffusion}, and \ref{fig:compresults_boundary}.

Figure \ref{fig:compresults_cells} gives the bone cell counts and bone mass percentage for the simulated bone remodeling event.  
Figure \ref{fig:compresults_cells}(a) shows the dynamics of the stromal cell population at position $x=0$.  The dynamics of this population at all other positions are similar to those shown in Figure \ref{fig:compresults_cells}(a).  Throughout the remodeling event, we see an increase in the number of stromal cells.
figure \ref{fig:compresults_cells}(b) shows the dynamics of the multiple myeloma tumor cell population.  For the first fifty days of the bone remodeling event, there is no significant change in the number of tumor cells.  However, in the last twenty-five days of the event we see an increase in the tumor cell population due to the multiple myeloma ``vicious cycle.''
Figures \ref{fig:compresults_cells}(c) and \ref{fig:compresults_cells}(d) show the dynamics of the osteoclast precursor and mature osteoclast cell populations.  Both populations decrease as the remodeling event continues.
Figures \ref{fig:compresults_cells}(e)  and \ref{fig:compresults_cells}(f) show the dynamics of the osteoblast precursor and mature osteoblast cell populations.  The osteoblast precursor population decreases in size quickly as osteoblast precursors are recruited to the mature osteoblast population.
Figure \ref{fig:compresults_cells}(g) shows the dynamics of the osteocyte population.  There is an initial decrease in the number of osteocytes due to the initiation of the bone remodeling event.  However, as the event continues, the number of osteocytes begins to increase due to the creation of new bone.
Figure \ref{fig:compresults_cells}(h) shows the percentage of bone mass throughout the bone remodeling event.  As the remodeling event progresses and the tumor cell population grows, the bone mass percentage decreases.

Figure \ref{fig:compresults_diffusion} shows the diffusion of ligands across the marrow.  Figure \ref{fig:compresults_diffusion}(a) shows the diffusion of BAFF and APRIL from the osteoclasts to the tumor cells.  Figure \ref{fig:compresults_diffusion}(b) shows the diffusion of MIP-1$\alpha$, IL-3, and TNF$\alpha$ from the tumor cells to the osteoclasts.  Figure \ref{fig:compresults_diffusion}(c) shows the diffusion of DKK1, IL-3, sclerotin, and sFRPs from the tumor cells to the osteoblasts.  Figure \ref{fig:compresults_diffusion}(d) shows the diffusion of IL-6, RANKL, GFs, and Activin A from the stromal cells to the osteoclasts.  Figure \ref{fig:compresults_diffusion}(e) shows the diffusion of Activin A from the stromal cells to the osteoblasts.

Figure \ref{fig:compresults_boundary} shows the movement of the bone/marrow interface and the marrow/tumor interface.  At time $t=0$, the bone/marrow interface is at $x=-1$ and the marrow/tumor interface is at $x=-1$.  As time progresses, the bone recedes and the tumor grows.  At time $t=75$, the bone/marrow interface is at $x=-1.2999$ and the marrow/tumor interface is at $x=0.5820$.




\section*{Acknowledgements}
We thank Dr. James Martin for calculating the diffusion values in Table \ref{tab:diffusion}.

\bibliographystyle{spbasic}      
\bibliography{library}   

\begin{thebibliography}{43}
\providecommand{\natexlab}[1]{#1}
\providecommand{\url}[1]{{#1}}
\providecommand{\urlprefix}{URL }
\expandafter\ifx\csname urlstyle\endcsname\relax
  \providecommand{\doi}[1]{DOI~\discretionary{}{}{}#1}\else
  \providecommand{\doi}{DOI~\discretionary{}{}{}\begingroup
  \urlstyle{rm}\Url}\fi
\providecommand{\eprint}[2][]{\url{#2}}

\bibitem[{Abe et~al(2004)Abe, Hiura, Wilde, Shioyasono, Moriyama, Hashimoto,
  Kido, Oshima, Shibata, Ozaki, Inoue, and Matsumoto}]{Abe2004}
Abe M, Hiura K, Wilde J, Shioyasono A, Moriyama K, Hashimoto T, Kido S, Oshima
  T, Shibata H, Ozaki S, Inoue D, Matsumoto T (2004) {Osteoclasts enhance
  myeloma cell growth and survival via cell-cell contact: a vicious cycle
  between bone destruction and myeloma expansion.} Blood 104(8):2484--91,
  \doi{10.1182/blood-2003-11-3839},
  \urlprefix\url{http://www.ncbi.nlm.nih.gov/pubmed/15187021}

\bibitem[{Abe et~al(2006)Abe, Kido, Hiasa, Nakano, Oda, Amou, and
  Matsumoto}]{Abe2006}
Abe M, Kido S, Hiasa M, Nakano A, Oda A, Amou H, Matsumoto T (2006) {BAFF and
  APRIL as osteoclast-derived survival factors for myeloma cells: a rationale
  for TACI-Fc treatment in patients with multiple myeloma.} Leukemia
  20(7):1313--5, \doi{10.1038/sj.leu.2404228},
  \urlprefix\url{http://www.nature.com.proxy.lib.uiowa.edu/leu/journal/v20/n7/full/2404228a.html}

\bibitem[{{American Cancer Society}(2015)}]{AmericanCancerSociety}
{American Cancer Society} (2015) {Multiple Myeloma}.
  http://www.cancer.org/cancer/multiplemyeloma/index,
  \urlprefix\url{http://www.cancer.org/cancer/multiplemyeloma/index}

\bibitem[{Ayati et~al(2010)Ayati, Edwards, Webb, and Wikswo}]{ayati2010}
Ayati BP, Edwards CM, Webb GF, Wikswo JP (2010) {A mathematical model of bone
  remodeling dynamics for normal bone cell populations and myeloma bone
  disease.} Biology direct 5(1):28, \doi{10.1186/1745-6150-5-28},
  \urlprefix\url{http://www.biology-direct.com/content/5/1/28}

\bibitem[{Bataille et~al(1990)Bataille, Delmas, Chappard, and
  Sany}]{Bataille1990}
Bataille R, Delmas PD, Chappard D, Sany J (1990) {Abnormal serum bone Gla
  protein levels in multiple myeloma. Crucial role of bone formation and
  prognostic implications.} Cancer 66(1):167--72,
  \urlprefix\url{http://www.ncbi.nlm.nih.gov/pubmed/2354403}

\bibitem[{Bataille et~al(1991)Bataille, Chappard, Marcelli, Dessauw, Baldet,
  Sany, and Alexandre}]{Bataille1991}
Bataille R, Chappard D, Marcelli C, Dessauw P, Baldet P, Sany J, Alexandre C
  (1991) {Recruitment of New Osteoblasts and Osteoclasts Is the Earliest
  Critical Event in the Pathogenesis of Human Multiple Myeloma}. Journal of
  Clinical Investigation 88(19):62--66, \doi{10.1172/JCI115305}

\bibitem[{Bellido et~al(2014)Bellido, Plotkin, and Bruzzaniti}]{Bellido2014}
Bellido T, Plotkin LI, Bruzzaniti A (2014) {Bone Cells}. In: Burr DB, Allen MR
  (eds) Basic and Applied Bone Biology, vol~14, Academic Press, chap~2, pp
  27--45, \doi{10.1016/B978-0-12-416015-6.00002-2}

\bibitem[{Bonewald(2011)}]{bonewald2011}
Bonewald LF (2011) {The amazing osteocyte}. Journal of Bone and Mineral
  Research 26(2):229--238, \doi{10.1002/jbmr.320}

\bibitem[{Bonewald and Dallas(1994)}]{Bonewald1994}
Bonewald LF, Dallas SL (1994) {Role of Active and latent Transforming Growth
  Factor in Bone Formation}. Journal of Cellular Biochemistry 55:350--357

\bibitem[{Buenzli(2015)}]{buenzli2015}
Buenzli PR (2015) {Osteocytes as a record of bone formation dynamics: A
  mathematical model of osteocyte generation in bone matrix}. Journal of
  Theoretical Biology 364:418--427, \doi{10.1016/j.jtbi.2014.09.028},
  \urlprefix\url{http://www.scopus.com/inward/record.url?eid=2-s2.0-84910685875{\&}partnerID=40{\&}md5=6ab8e5b855fa2a6cd8459ed392b61b4c}

\bibitem[{Burr(2002)}]{Burr2002}
Burr DB (2002) {Targeted and Nontargeted Remodeling}. Bone 30(1):2--4,
  \urlprefix\url{http://www.researchgate.net/profile/David{\_}Burr3/publication/11565284{\_}Targeted{\_}and{\_}nontargeted{\_}remodeling/links/54b573370cf26833efd26c95.pdf}

\bibitem[{Colucci et~al(2011)Colucci, Brunetti, Oranger, Mori, Sardone,
  Specchia, Rinaldi, Curci, Liso, Passeri, Zallone, Rizzi, and
  Grano}]{Colucci2011}
Colucci S, Brunetti G, Oranger A, Mori G, Sardone F, Specchia G, Rinaldi E,
  Curci P, Liso V, Passeri G, Zallone A, Rizzi R, Grano M (2011) {Myeloma cells
  suppress osteoblasts through sclerostin secretion.} Blood cancer journal
  1(6):e27, \doi{10.1038/bcj.2011.22},
  \urlprefix\url{http://www.pubmedcentral.nih.gov/articlerender.fcgi?artid=3255263{\&}tool=pmcentrez{\&}rendertype=abstract}

\bibitem[{Drake(2014)}]{drake2014}
Drake MT (2014) {Myeloma Bone Disease}. In: Gertz MA, Rajkumar SV (eds)
  Multiple Myeloma, Springer, chap~17, \doi{10.1007/978-1-4614-8520-9},
  \urlprefix\url{http://link.springer.com/10.1007/978-1-4614-8520-9}

\bibitem[{Ehrlich and Roodman(2005)}]{Ehrlich2005}
Ehrlich LA, Roodman GD (2005) {The role of immune cells and inflammatory
  cytokines in Paget's disease and multiple myeloma}. Immunological Reviews
  208:252--266, \doi{IMR323 [pii] 10.1111/j.0105-2896.2005.00323.x}

\bibitem[{Eudy et~al(2015)Eudy, Gastonguay, Baron, and Riggs}]{Eudy2015}
Eudy RJ, Gastonguay MR, Baron KT, Riggs M (2015) {Connecting the Dots: Linking
  Osteocyte Activity and Therapeutic Modulation of Sclerostin by Extending a
  Multiscale Systems Model}. CPT: Pharmacometrics {\&} Systems Pharmacology
  (May), \doi{10.1002/psp4.12013}

\bibitem[{Evans et~al(1989)Evans, Galasko, and Ward}]{Evans1989}
Evans CE, Galasko CSB, Ward C (1989) {Does Myeloma Secrete and Osteoblast
  Inhibiting Factor?} The Journal of Bone and Joint Surgery 71(2):288--290

\bibitem[{Giuliani et~al(2012)Giuliani, Ferretti, Bolzoni, Storti, Lazzaretti,
  {Dalla Palma}, Bonomini, Martella, Agnelli, Neri, Ceccarelli, and
  Palumbo}]{Giuliani2012}
Giuliani N, Ferretti M, Bolzoni M, Storti P, Lazzaretti M, {Dalla Palma} B,
  Bonomini S, Martella E, Agnelli L, Neri A, Ceccarelli F, Palumbo C (2012)
  {Increased osteocyte death in multiple myeloma patients: role in
  myeloma-induced osteoclast formation.} Leukemia 26(6):1391--401,
  \doi{10.1038/leu.2011.381},
  \urlprefix\url{http://www.nature.com.proxy.lib.uiowa.edu/leu/journal/v26/n6/full/leu2011381a.html}

\bibitem[{Graham et~al(2012)Graham, Ayati, Ramakrishnan, and
  Martin}]{Graham2012}
Graham JM, Ayati BP, Ramakrishnan PS, Martin JA (2012) {Towards a new spatial
  representation of bone remodeling.} Mathematical biosciences and engineering
  : MBE 9(2):281--95,
  \urlprefix\url{http://www.pubmedcentral.nih.gov/articlerender.fcgi?artid=3708700{\&}tool=pmcentrez{\&}rendertype=abstract}

\bibitem[{Graham et~al(2013)Graham, Ayati, Holstein, and Martin}]{graham2013}
Graham JM, Ayati BP, Holstein SA, Martin JA (2013) {The role of osteocytes in
  targeted bone remodeling: a mathematical model.} PloS one 8(5):e63,884,
  \doi{10.1371/journal.pone.0063884},
  \urlprefix\url{http://journals.plos.org/plosone/article?id=10.1371/journal.pone.0063884}

\bibitem[{Ji et~al(2015)Ji, Genever, and Fagan}]{Ji2015}
Ji B, Genever PG, Fagan MJ (2015) {A virtual approach to evaluate therapies for
  management of multiple myeloma induced bone disease}. International Journal
  for Numerical Methods in Biomedical Engineering e02735:1--18,
  \doi{10.1002/cnm}

\bibitem[{Komarova et~al(2003)Komarova, Smith, Dixon, Sims, and
  Wahl}]{komarova2003}
Komarova SV, Smith RJ, Dixon S, Sims SM, Wahl LM (2003) {Mathematical model
  predicts a critical role for osteoclast autocrine regulation in the control
  of bone remodeling}. Bone 33(2):206--215,
  \doi{10.1016/S8756-3282(03)00157-1},
  \urlprefix\url{http://www.sciencedirect.com/science/article/pii/S8756328203001571}

\bibitem[{Kular et~al(2012)Kular, Tickner, Chim, and Xu}]{Kular2012}
Kular J, Tickner J, Chim SM, Xu J (2012) {An overview of the regulation of bone
  remodelling at the cellular level.} Clinical biochemistry 45(12):863--73,
  \doi{10.1016/j.clinbiochem.2012.03.021},
  \urlprefix\url{http://www.sciencedirect.com/science/article/pii/S000991201200149X}

\bibitem[{Michigami et~al(2000)Michigami, Shimizu, Williams, Niewolna, Dallas,
  Mundy, and Yoneda}]{Michigami2000}
Michigami T, Shimizu N, Williams PJ, Niewolna M, Dallas SL, Mundy GR, Yoneda T
  (2000) {Cell-cell contact between marrow stromal cells and myeloma cells via
  VCAM-1 and alpha(4)beta(1)-integrin enhances production of
  osteoclast-stimulating activity.} Blood 96(5):1953--60,
  \urlprefix\url{http://www.bloodjournal.org/content/96/5/1953.abstract}

\bibitem[{Mori et~al(2004)Mori, Shimizu, Dallas, Niewolna, Story, Williams,
  Mundy, and Yoneda}]{Mori2004}
Mori Y, Shimizu N, Dallas M, Niewolna M, Story B, Williams PJ, Mundy GR, Yoneda
  T (2004) {Anti-alpha4 integrin antibody suppresses the development of
  multiple myeloma and associated osteoclastic osteolysis.} Blood
  104(7):2149--54, \doi{10.1182/blood-2004-01-0236},
  \urlprefix\url{http://www.ncbi.nlm.nih.gov/pubmed/15138161}

\bibitem[{Mundy et~al(1974)Mundy, Raisz, Cooper, Schechter, and
  Salmon}]{Mundy1974}
Mundy GR, Raisz LG, Cooper RA, Schechter GP, Salmon SE (1974) {Evidence for the
  Secretion of an Osteoclast Stimulating Factor in Myeloma}. The New England
  Journal of Medicine 291(20):1041--1046,
  \urlprefix\url{http://www.nejm.org.proxy.lib.uiowa.edu/doi/full/10.1056/NEJM197411142912001}

\bibitem[{Neve et~al(2012)Neve, Corrado, and Cantatore}]{neve2012}
Neve A, Corrado A, Cantatore FP (2012) {Osteocytes: central conductors of bone
  biology in normal and pathological conditions.} Acta physiologica (Oxford,
  England) 204(3):317--30, \doi{10.1111/j.1748-1716.2011.02385.x},
  \urlprefix\url{http://www.ncbi.nlm.nih.gov/pubmed/22099166}

\bibitem[{Oshima et~al(2005)Oshima, Abe, Asano, Hara, Kitazoe, Sekimoto,
  Tanaka, Shibata, Hashimoto, Ozaki, Kido, Inoue, and Matsumoto}]{Oshima2005}
Oshima T, Abe M, Asano J, Hara T, Kitazoe K, Sekimoto E, Tanaka Y, Shibata H,
  Hashimoto T, Ozaki S, Kido S, Inoue D, Matsumoto T (2005) {Myeloma cells
  suppress bone formation by secreting a soluble Wnt inhibitor, sFRP-2}. Blood
  106(9):3160--3165, \doi{10.1182/blood-2004-12-4940}

\bibitem[{Parfitt(2002)}]{Parfitt2002}
Parfitt A (2002) {Targeted and nontargeted bone remodeling: relationship to
  basic multicellular unit origination and progression}. Bone 30(1):5--7,
  \doi{10.1016/S8756-3282(01)00642-1},
  \urlprefix\url{http://www.sciencedirect.com/science/article/pii/S8756328201006421}

\bibitem[{Parfitt(1994)}]{Parfitt1994}
Parfitt AM (1994) {Osteonal and hemi-osteonal remodeling: the spatial and
  temporal framework for signal traffic in adult human bone.} Journal of
  cellular biochemistry 55(3):273--86, \doi{10.1002/jcb.240550303},
  \urlprefix\url{http://www.ncbi.nlm.nih.gov/pubmed/7962158}

\bibitem[{Plotkin and Bivi(2014)}]{Plotkin2014}
Plotkin LI, Bivi N (2014) {Local Regulation of Bone Cell Function}. In: Burr
  DB, Allen MR (eds) Basic and Applied Bone Biology, Academic Press, chap~3, pp
  47--74, \doi{10.1016/B978-0-12-416015-6.00003-4},
  \urlprefix\url{http://dx.doi.org/10.1016/B978-0-12-416015-6.00003-4}

\bibitem[{Ryser et~al(2009)Ryser, Nigam, and Komarova}]{Ryser2009}
Ryser MD, Nigam N, Komarova SV (2009) {Mathematical Modeling of Spatio-Temporal
  Dynamics of a Single Bone Multicellular Unit}. Journal of Bone and Mineral
  Research 24(5):860--870, \doi{10.1359/jbmr.081229}

\bibitem[{Ryser et~al(2010)Ryser, Komarova, and Nigam}]{Ryser2010}
Ryser MD, Komarova SV, Nigam N (2010) {The Cellular Dynamics of Bone
  Remodeling: A Mathematical Model}. SIAM J Appl Math 70(6):1899--1921

\bibitem[{Savageau(1969{\natexlab{a}})}]{savageau1969}
Savageau MA (1969{\natexlab{a}}) {Biochemical systems analysis: I. Some
  mathematical properties of the rate law for the component enzymatic
  reactions}. Journal of Theoretical Biology 25(3):365--369,
  \doi{10.1016/S0022-5193(69)80026-3},
  \urlprefix\url{http://www.sciencedirect.com/science/article/pii/S0022519369800263}

\bibitem[{Savageau(1969{\natexlab{b}})}]{savageau1969a}
Savageau MA (1969{\natexlab{b}}) {Biochemical systems analysis: II. The
  steady-state solutions for an n-pool system using a power-law approximation}.
  Journal of Theoretical Biology 25(3):370--379,
  \doi{10.1016/S0022-5193(69)80027-5},
  \urlprefix\url{http://www.sciencedirect.com/science/article/pii/S0022519369800275}

\bibitem[{Savageau(1970)}]{savageau1970}
Savageau MA (1970) {Biochemical systems analysis: III. Dynamic solutions using
  a power-law approximation}. Journal of Theoretical Biology 26(2):215--226,
  \doi{10.1016/S0022-5193(70)80013-3},
  \urlprefix\url{http://www.sciencedirect.com/science/article/pii/S0022519370800133}

\bibitem[{Savageau(1976)}]{savageau1976}
Savageau MA (1976) {Biochemical systems analysis A study of function and design
  in molecular biology}. Addison-Wesley, Reading, MA

\bibitem[{Silbermann and Roodman(2013)}]{Silbermann2013}
Silbermann R, Roodman GD (2013) {Myeloma bone disease: Pathophysiology and
  management}. Journal of Bone Oncology \doi{10.1016/j.jbo.2013.04.001}

\bibitem[{Tian et~al(2003)Tian, Zhan, Walker, Rasmussen, Ma, Barlogie, and
  Shaughnessy}]{Tian2003}
Tian E, Zhan F, Walker R, Rasmussen E, Ma Y, Barlogie B, Shaughnessy JD (2003)
  {The role of the Wnt-signaling antagonist DKK1 in the development of
  osteolytic lesions in multiple myeloma.} The New England journal of medicine
  349(26):2483--94, \doi{10.1056/NEJMoa030847},
  \urlprefix\url{http://www.ncbi.nlm.nih.gov/pubmed/14695408}

\bibitem[{Valentin-Opran et~al(1982)Valentin-Opran, Charhon, Meunier, Edouard,
  and Arlot}]{Valentin-Opran1982}
Valentin-Opran A, Charhon SA, Meunier PJ, Edouard CM, Arlot ME (1982)
  {Quantitative histology of myeloma-induced bone changes}. British journal of
  haematology 52(4):601--10,
  \urlprefix\url{http://www.ncbi.nlm.nih.gov/pubmed/7138789}

\bibitem[{Vallet et~al(2010)Vallet, Mukherjee, Vaghela, Hideshima, Fulciniti,
  Pozzi, Santo, Cirstea, Patel, Sohani, Guimaraes, Xie, Chauhan, Schoonmaker,
  Attar, Churchill, Weller, Munshi, Seehra, Weissleder, Anderson, Scadden, and
  Raje}]{Vallet2010}
Vallet S, Mukherjee S, Vaghela N, Hideshima T, Fulciniti M, Pozzi S, Santo L,
  Cirstea D, Patel K, Sohani AR, Guimaraes A, Xie W, Chauhan D, Schoonmaker JA,
  Attar E, Churchill M, Weller E, Munshi N, Seehra JS, Weissleder R, Anderson
  KC, Scadden DT, Raje N (2010) {Activin A promotes multiple myeloma-induced
  osteolysis and is a promising target for myeloma bone disease.} Proceedings
  of the National Academy of Sciences of the United States of America
  107(11):5124--9, \doi{10.1073/pnas.0911929107},
  \urlprefix\url{http://www.pubmedcentral.nih.gov/articlerender.fcgi?artid=2841922{\&}tool=pmcentrez{\&}rendertype=abstract}

\bibitem[{Voit(2000)}]{voit2000}
Voit EO (2000) {Computational Analysis of Biochemical Systems}. Cambridge
  University Press, Cambridge, UK

\bibitem[{Wahlin et~al(2009)Wahlin, Holm, Osterman, and Norberg}]{Wahlin2009}
Wahlin A, Holm J, Osterman G, Norberg B (2009) {Evaluation of Serial Bone X-ray
  Examination in Multiple Myeloma}. Acta Medica Scandinavica 212(6):385--387,
  \doi{10.1111/j.0954-6820.1982.tb03234.x},
  \urlprefix\url{http://doi.wiley.com/10.1111/j.0954-6820.1982.tb03234.x}

\bibitem[{Wang et~al(2011)Wang, Pivonka, Buenzli, Smith, and
  Dunstan}]{Wang2011}
Wang Y, Pivonka P, Buenzli PR, Smith DW, Dunstan CR (2011) {Computational
  modeling of interactions between multiple myeloma and the bone
  microenvironment.} PloS one 6(11):e27,494,
  \doi{10.1371/journal.pone.0027494},
  \urlprefix\url{http://journals.plos.org/plosone/article?id=10.1371/journal.pone.0027494}

\end{thebibliography}

\newpage


\begin{figure}
\begin{center}\includegraphics[width=0.75\textwidth]{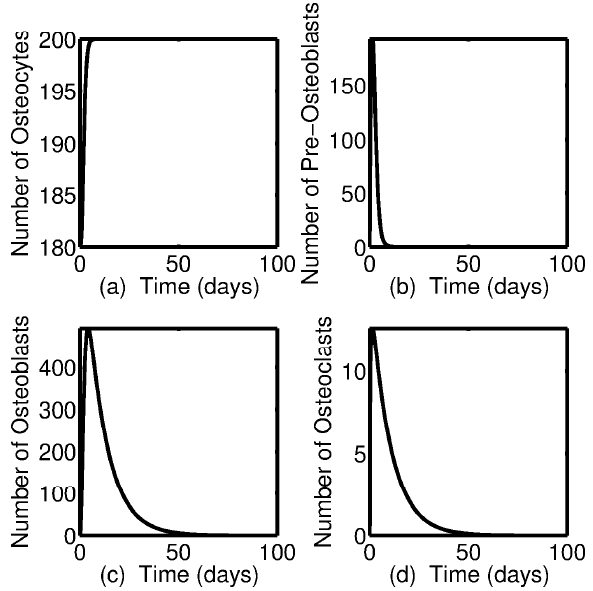} \end{center}
\caption{Population dynamics during a bone remodeling event (without the presence of a tumor), as simulated by \cite{graham2013}}\end{figure}

\begin{figure}
\begin{center}\includegraphics[width=0.75\textwidth]{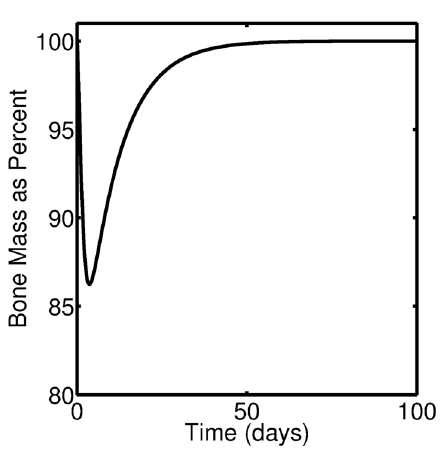} \end{center}
\caption{Bone volume dynamics during a bone remodeling event (without the presence of a tumor), as simulated by  \cite{graham2013}}\end{figure}

\begin{table}
  \rowcolors{2}{gray!25}{white}
  \begin{center}
  \begin{tabular}{l|l}
    \rowcolor{gray!50}
    Symbol & Definition \\
    $C(t)$ & Number of osteoclasts at time $t$ \\
    $B(t)$ & Number of osteoblasts at time $t$ \\
    $\overline{C}$ & Number of osteoclasts in the steady-state \\
    $\overline{B}$ & Number of osteoblasts in the steady-state \\
    $z(t)$ & Percentage of bone mass at time $t$ \\
    $T(t)$ & Number of tumor cells at time $t$ \\
    $L_{T}$ & Maximum tumor size \\
    $\gamma_{T}$ & Tumor growth constant \\
    $V_{1}(t), V_{2}(t)$ & Treatment functions
  \end{tabular}
  \end{center}
\caption{Definitions of symbols used in Section \ref{sec:0dim}}
\label{tab:variables1}
\end{table}

\begin{table}
  \rowcolors{2}{gray!25}{white}
  \begin{center}
  \begin{tabular}{l|l}
    \rowcolor{gray!50}
    Symbol & Definition \\
    $Y(t)$ & Number of osteocytes at time $t$ \\
    $B_{P}(t)$ & Number of osteoblast precursors at time $t$ \\
    $B(t)$ & Number of osteoblasts at time $t$ \\
    $C(t)$ & Number of osteoclasts at time $t$ \\
    $z(t)$ & Percentage of bone mass at time $t$ \\
    $K_{Y}$ & Osteocyte population threshold for sclerostin production \\
  \end{tabular}
  \end{center}
\caption{Definitions of symbols used in Section \ref{sec:jason}}
\label{tab:variables2}
\end{table}

\begin{table}
  \rowcolors{2}{gray!25}{white}
  \begin{center}
  \begin{tabular}{l|l}
    \rowcolor{gray!50}
    Symbol & Definition \\
    $S(t)$ & Number of stromal cells at time $t$\\
    $T(t)$ & Number of tumor cells at time $t$\\
    $C_{P}(t)$ & Number of osteoclast precursors at time $t$\\
    $C(t)$ & Number of osteoclasts at time $t$ \\
    $B_{P}(t)$ & Number of osteoblast precursors at time $t$ \\
    $B(t)$ & Number of osteoblasts at time $t$ \\
    $Y(t)$ & Number of osteocytes at time $t$ \\
    $L_{C}(t)$ & BAFF and APRIL, diffusing from the osteoclasts to the tumor cells\\
    $L_{T_{1}}(t)$ & MIP-1$\alpha$, IL-3, and TNF$\alpha$, diffusing from the tumor cells to the osteoclasts\\
    $L_{T_{2}}(t)$ & DKK1, IL-3, sclerostin, and sFRPs, diffusing from the tumor cells to the osteoblasts\\
    $L_{S_{1}}(t)$ & IL-6, RANKL, GFs, and Activin A, diffusing from the stromal cells to the osteoclasts\\
    $L_{S_{2}}(t)$ & Activin A, diffusing from the stromal cells to the osteoblasts\\
    $z(t)$ & Percentage of bone mass at time $t$ \\
    $\ell(t)$ & Position of the bone/marrow interface at time $t$\\
    $r(t)$ & Position of the marrow/tumor interface at time $t$ \\
    $K_{Y}$ & Osteocyte population threshold for sclerostin production \\
  \end{tabular}
  \end{center}
\caption{Definitions of symbols used in Section \ref{sec:catie}}
\label{tab:variables3}
\end{table}

\begin{table}
  \rowcolors{2}{gray!25}{white}
  \begin{center}
  \begin{tabular}{l|l||l|l}
    \rowcolor{gray!50}
    Parameter & Value & Parameter & Value \\
    $\alpha_{1}$ & 0.005 & $K_{Y}$ & 10\\
    $\alpha_{2}$ & 0.0003 & $\epsilon$ & 1\\
    $\alpha_{3}$ & 0.0001 & $g_{11}$ & 1\\
    $\alpha_{4}$ & 0.01 & $g_{12}$ & 1\\
    $\alpha_{5}$ & 0.01 & $g_{21}$ & 1\\
    $\alpha_{6}$ & 0.01 & $g_{22}$ & 1\\
    $\beta_{1}$ & 0.01 & $g_{31}$ & 1\\
    $\beta_{2}$ & 0.008 &  $g_{32}$ & 1\\
    $\beta_{3}$ & 0.01 & $g_{41}$ & 1\\
    $\beta_{4}$ & 0.01 & $g_{51}$ & 1\\
    $\beta_{5}$ & 0.01 & $g_{52}$ & 1\\
    $\beta_{6}$ & 0.01 & $h_{11}$ & -1\\
    $\beta_{7}$ & 0.1 & $h_{12}$ & 1\\
    $\gamma_{1}$ & 0.01 & $h_{13}$ & 1\\
    $\gamma_{2}$ & 38.4 & $h_{14}$ & 1\\
    $\gamma_{3}$ & 0.00390625 & $h_{15}$ & 1\\
    $\delta_{11}$ & 0.1037 & $h_{21}$ & 1\\
    $\delta_{12}$ & 0.01 & $h_{22}$ & -0.8\\
    $\delta_{21}$ & 0.1210 & $h_{23}$ & -0.8\\
    $\delta_{22}$ & 0.01 & $h_{24}$ & -0.8\\
    $\delta_{31}$ & 0.1037 & $h_{25}$ & -0.8\\
    $\delta_{32}$ & 0.01 &  $h_{26}$ & 1\\
    $\delta_{41}$ & 0.1728 & $h_{31}$ & 1\\
    $\delta_{42}$ & 0.01 & $h_{32}$ & 0.5\\
    $\delta_{51}$ & 0.1063 & $ f_{21}$ & 0.65\\
    $\delta_{52}$ & 0.01 & $a$ & 0.01\\
    $k_{1}$ & 0.1 &    $b$ & -0.03\\
    $k_{2}$ & 0.01 & &
  \end{tabular}
  \end{center}
\caption{Parameter values used in Section \ref{sec:catie}}
\label{tab:parameters3}
\end{table}

\begin{table}
\rowcolors{2}{gray!25}{white}
\begin{center}
\begin{tabular}{|l|l|l|l|l|}\hline
\rowcolor{gray!50}
  \multicolumn{1}{|p{1.7cm}|}{\centering \hspace{0.1cm} \\ \textbf{Peptide} \\  \ \\ \ \\ \hspace{0.1cm}} & \multicolumn{1}{|p{1.7cm}|}{\centering \hspace{0.1cm} \\ \textbf{Molecular} \\  \textbf{Weight} \\ {\footnotesize(kDa)} \\ \hspace{0.1cm}} & \multicolumn{1}{|p{2cm}|}{\centering \hspace{0.1cm} \\ \textbf{Diffuusion Coefficient} \\ \textbf{(Literature)}\\ {\footnotesize(cm$^{2}$/sec)} \\ \hspace{0.1cm}} & \multicolumn{1}{|p{1.3cm}|}{\centering  \hspace{0.1cm} \\ \textbf{Stokes} \\  \textbf{Radius} \\ {\footnotesize(nm)} \\ \hspace{0.1cm}} & \multicolumn{1}{|p{2.2cm}|}{\centering \hspace{0.1cm} \\ \textbf{Diffusion Coefficient (Calculated)} \\ \ \\ {\footnotesize(cm$^{2}$/sec)} \\ \hspace{0.1cm}} \\\hline
  IL-2 & $15.1$ & $1.0 \times 10^{-6}$ & $1.602$ & $1.36 \times 10^{-6}$ \\
  IL-3 & $16.2$ & $1.0 \times 10^{-6}$ & $1.6328$ & $1.35 \times 10^{-6}$ \\
  IL-6 & $23.7$ & $9.0 \times 10^{-7}$ & $1.9628$ & $1.21 \times 10^{-6}$ \\
  RANKL & $35$ & & $2.073$ & $1.17 \times 10^{-6}$ \\
  OPG & $60$ & & $2.463$ & $1.01 \times 10^{-6}$ \\
  BAFF & $31$ & & $2.0106$ & $1.20 \times 10^{-6}$ \\
  APRIL & $28$ & & $1.9638$ & $1.21 \times 10^{-6}$ \\
  MIP-1$\alpha$ & $8$ & & $1.272$ & $1.49 \times 10^{-6}$ \\
  TNF & $17$ & & $1.668$ & $1.33 \times 10^{-6}$ \\
  DKK-1 & $26$ & & $1.9326$ & $1.23 \times 10^{-6}$ \\
  Sclerostin & $23$ & $9.0 \times 10^{-7}$ & $1.932$ & $1.23 \times 10^{-6}$ \\
  sFRP-1 & $33$ & & $2.0418$ & $1.18 \times 10^{-6}$ \\
  GFs & & & & $2.00 \times 10^{-6}$ \\
  Activin A & $26.2$ & & $1.93572$ & $1.23 \times 10^{-6}$ \\
  Hemoglobin & $68$ & $6.9 \times 10^{-7}$ & & \\
  \hline
  \end{tabular}
  \end{center}
  \caption{Table of known and calculated diffusion coefficients.}
  \label{tab:diffusion}
  \end{table}

\end{document}